# On the propagation speed of wavy metric tensors

A. LOINGER

Dipartimento di Fisica, Università di Milano

Via Celoria, 16 − 20133 Milano, Italy

**Summary.** − The undulating metric tensors of general relativity do not possess a special and common velocity of propagation. Indeed, their velocity can even coincide with the speed of thought.

PACS. 04.30 − Gravitational waves and radiation: theory.

**1**. − According to a rooted belief, any exact (nonperturbative) solution to the Einstein equations with a mass tensor equal to zero

$$(1.1) \qquad R_{jk} = 0, \quad (j, k = 0, 1, 2, 3) \,,$$

having an undulatory character − and a curvature tensor different from zero − is propagated with the *fundamental velocity* [1]. I affirm that this conviction is erroneous. In general relativity the fundamental velocity is only a universal constant, on the same footing of the gravitational constant. People have been misled by a superficial analogy with Maxwell theory in *Minkowski* spacetime; however, this analogy has some meaning only for the linearized and perturbative formulations of Einstein theory, which postulate a *background* spacetime.

*In primis*, and generally speaking, a conceptual pilaster of general relativity is just the absence of a privileged velocity of the physical phenomena. And so far as the wavy solutions of eqs. (1.1) are concerned, the belief in question is **not** supported by the theory of the characteristics of Einstein equations, which was created by Levi-Civita in 1930 [2]. It is, simply, an unjustified assumption.

arXiv: gr-qc/0007048    July 19th, 2000

Of course, it is possible to *select* undulating metric tensors, solutions of eqs. (1.1), which are propagated − in well-characterized reference systems − with the fundamental velocity [3], but we are always free to choose other − and equally legitimate − co-ordinate frames for which the propagation happens with *ad libitum* velocities, from zero to the speed of thought. Further, the undulatory characters themselves can be destroyed by suitable changes of the reference systems.

It is evident that any undulation of this kind − which, moreover, has only a ***pseudo*** energy, and therefore ***cannot effect an actual geodesic deviation*** − does ***not*** represent a physical wave: it is only a formal expression, in spite of the fact that its Riemann tensor $R_{jklm}$ is different from zero.

Remark that in general relativity the very concept of a $g_{jk}$-propagation (and of a $R_{jklm}$-propagation) is rather peculiar, and must be properly understood. To be determinate, consider a fictive universe only composed of a wavy metric field $g_{jk}$, solution of eqs. (1.1). We can say that such a world is generated by this $g_{jk}$. It would be a nonsense to think of our "gravitational" field as an undulation which is propagated in a *pre-existent* spacetime, because in general relativity spacetime coincides always with the collection of the world points for which the $g_{jk}$'s have some values. Indeed, in the exact (nonlinearized and nonperturbative) theory there is *no* pre-existent spacetime, not even in the simple form of a topological manifold.

In a series of papers [4], I have developed many arguments demonstrating the nonexistence of the gravitational waves. The nonexistence of a privileged velocity for the undulatory metric fields is another and straightforward proof that the notion of gravity radiation belongs to the realms of science fiction, together with the notion of black hole.

**2**. − The superficial analogy between Maxwell field and Einstein field is rather deceitful. Friedmann's cosmological models yield a simple and remarkable demonstration of this fact.

Indeed, these models represent *rigorous* solutions to Einstein equations for the problem of purely gravitational motions of a "dust" characterized by the mass tensor $T^{jk} = \rho \left( dx^j / ds \right)\left( dx^k / ds \right)$, where $\rho$ is the invariant density. Now, ***no*** gravitational radiation appears in Friedmann's results. On the contrary, in the analogous problems of the motions of electrically charged "dusts" in Minkowski spacetime the emission of electromagnetic waves is unavoidable.

Friedmann models give a further proof of the nonexistence of the gravitational waves.

**3**. − ***A question***: What is the behaviour of $g_{jk}$ when, for instance, a *supernova* explodes? ***Answer***: A sufficiently near observer would register a sudden variation of Einstein field, *qualitatively* similar to the corresponding variation of Newton field.

> "Sie sagen: das mutet mich nicht an!
> Und meinen, sie hättens abgetan."
> J.W. v. Goethe